\documentclass[prd,amsmath,amssymb,twocolumn]{revtex4-1} %twocolumn, showpacs,preprintnumbers

\usepackage[utf8]{inputenc}  
\usepackage{graphicx}% Include figure files
\usepackage{bm}% bold math

\newcommand{\etal}{ \textit{et al.} }

\newcommand{\msun}{\,\mbox{M}_\odot}

\newcommand{\PR}{\mathcal P_{\mathcal R}}
\newcommand{\Pd}{\mathcal P_{\delta}}

\begin{document}

% \preprint{}

\title{
A large change in the predicted number of small halos due to a small amplitude oscillating inflaton potential.
% Increasing (or decreasing) the predicted number of small halos due to a small amplitude oscillating inflaton potential
}
\author{Luiz Felippe S. Rodrigues}
\email{felippe@astro.iag.usp.br}
\author{Reuven Opher}
\email{opher@astro.iag.usp.br}
\affiliation{Instituto de Astronomia Geofísica e Ciências Atmosféricas, Universidade de S\~{a}o Paulo\\
Rua do Mat\~{a}o 1226, Cidade Universit\'{a}ria, CEP 05508-900, S\~{a}o Paulo, SP, Brazil
}
\date{\today}

\begin{abstract}
A smooth inflaton potential is generally assumed when calculating the primordial power spectrum, implicitly assuming that a very small oscillation in the inflaton potential creates a negligible change in the predicted halo mass function. We show that this is not true. We find that a small oscillating perturbation in the inflaton potential in the slow-roll regime can alter significantly the predicted number of small halos. A class of models derived from supergravity theories gives rise to inflaton potentials with a large number of steps and many transplanckian effects may generate oscillations in the primordial power spectrum. The potentials we study are the simple quadratic (chaotic inflation) potential with superimposed small oscillations for small field values. Without leaving the slow-roll regime, we find that for a wide choice of parameters, the predicted number of halos change appreciably. For the oscillations beginning in the $10^7-10^8\msun$ range, for example, we find that only a 5\% change in the amplitude of the chaotic potential causes a 50\% suppression of the number of halos for masses between $10^7-10^8\msun$ and an increase in the number of halos for masses $< 10^6 \msun$ by factors $\sim 15-50$. We suggest that this might be a solution to the problem of the lack of observed dwarf galaxies in the range $10^7-10^8\msun$. This might also be a solution to the reionization problem where a very large number of Population \textsc{III} stars in low mass halos are required.
% 
% We show that small oscillating perturbations in the inflaton potential, in the slow-roll regime, can alter significantly the predicted number of small halos. The potentials  we study are the simple quadratic (chaotic inflation) potential with superimposed small oscillations for small field values. We find that for a wide choice of parameters it is possible to increase or decrease the predicted number of halos without leaving the slow-roll regime. We discuss the implications of these results on the missing satellites problem. We find, for example, that saw-tooth oscillations, with an amplitude of only 5\% of the chaotic potential, causes a $\sim 50\%$ suppression of the number of halos  for masses between $10^7$--$10^8\msun$ and an increase in the number for masses $10^5$--$10^6\msun$. Similar results are obtained for small amplitude sinusoidal oscillations.
\end{abstract}
\maketitle

\section{Introduction}

Recently inflation has become an essential part of our description of the universe. Not only  does it solve the classical cosmological problems of flatness, horizon and relics, but also provides precise predictions for the primordial density inhomogeneities, predictions that are in good agreement with existing observations\cite{WMAP5}.

Albeit these successes, the specific details of inflation are still unknown, since many different physical mechanisms and fields may generate a phase of accelerated cosmic expansion. For simplicity, it has become common practice to employ a scalar field, {the inflaton}, with a simple law for its potential, to generate inflation. This  picture is usually understood as the effective counterpart of a deeper -- and probably more complicated -- theory. In this context, the most generally employed inflationary theory is the so-called chaotic inflation, characterized by a simple quadratic potential with an initial  high field value for the inflaton.

Recently, due to advances in data quality and in anticipation of the data from the Planck satellite \cite{planck} and the Large Synoptic Survey Telescope (LSST) \cite{LSST}), for example, several groups started investigating more complicated forms for the inflaton potential to explain the present observational data. Pahud\etal  \cite{Pahud2008} used the CMB data to look for the presence of a general sinusoidal oscillation imprinted on the inflaton potential, for large field values (i.e., large spatial dimensions), placing strong limits on the amplitude of these osillations. Ichiki\etal \cite{Ichiki2009a} found -- with a 99,995\% confidence level -- an oscillatory modulation for large spatial dimensions at $k\simeq 0.009\text{ Mpc}^{-1}$, performing a Monte-Carlo Markov-Chain analysis using the CMB data, confirming similar results obained from the analysis of the CMB data using different techniques \cite{Ichiki2009a, Ichiki2009}.

From a theoretical perspective, features in the inflaton potential are well motivated.  Adams\etal \cite{Adams1997} showed that a class of models derived from supergravity theories gives rise to inflaton potentials with a large number of steps, each of these corresponding to a symmetry-breaking phase transition in a field coupled with the inflaton. Also, many transplanckian effects may generate oscillations in the primordial power spectrum of density inhomogeneities which could be described by an effective oscillating inflaton potential \cite{Brandenberger2000, Martin2003}.

A present major problem in astrophysics is the large discrepancy between the predicted and the observed number of dark matter halos of mass $\lesssim 10^8\msun$. N-body simulations designed to probe the formation and evolution of dark matter structures on small scales found $\sim 10^3$ dark matter halos with masses from $10^7\msun$ to $10^9\msun$ \cite{ViaLactea,Aquarius}. However, very much fewer  small galaxies of comparable masses are observed in the Local Group \cite{ViaLactea}.

Several solutions to this number discrepancy have been proposed, ranging from selection effects in the observations \cite{Koposov2008} to complex baryonic interactions that may have swept out the baryonic gas from the small halos \cite{Maccio2009,Koposov2009}. The former explanation for the discrepancy can  only dimish the problem but not resolve it: taking into account the limitations of present day observations, one may extrapolate the number of dwarf galaxies to -- at most -- a few hundred. The latter type of solution can only be tested with semi-analytical models (for a review see  \cite{Baugh2006}), which strongly rely on the details of the physics of star formation, which is not completely understood. The ejection of gas would leave thousands of empty small dark matter halos essentially intact. There is still no clear evidence of the presence of these objects in the dynamics of the Local Group.

In this work we propose an alternative approach. Bearing in mind the growing plausability of oscillating features in the inflaton potential, we examine to what extent a simple localized oscillating modification of the inflaton potential for small field values can change the number of small dark matter halos. We restrict our analysis to perturbations of the inflaton potential which are still in the slow-roll regime.

In sections \ref{secsr} and \ref{secmf} we review the formalism and the derivation of the important relations used. In section \ref{secpot} we discuss two simple modified oscillatory potentials: a saw-tooth modification and a sinusoidal modification. In section \ref{secresults} we present our results and in section \ref{secconcl} our conclusions.

\section{Slow-roll formalism}\label{secsr}
The primordial universe is assumed to be filled by an approximately homogeneous  scalar field, the inflaton, governed by a Klein-Gordon equation of motion
\begin{equation}
 \ddot \phi + 3 H\dot \phi + \frac{d V}{d \phi}=0 \label{KG}\quad\mbox{ . }
\end{equation}
The Friedmann equation becomes
\begin{equation}
 \left(\frac{\dot a}{a}\right)^2= \frac{8\pi}{3 \,m_{pl}^2} \left[ V(\phi) + \frac{1}{2} \dot \phi^2 \right]\mbox{ .}\label{Fried}
\end{equation}

In order to have an inflationary period, the second derivative of the field and the kinetic term of the Friedmann equation must both be small when compared with the other terms. This can be obtained under the conditions of slow-roll,
\begin{equation}
\epsilon \equiv \frac{1}{8}\left(\frac{V'(f)}{V(f)} \right)^2\ll1 \label{slowroll1}
\end{equation}
and
\begin{equation}
 \eta \equiv \frac{1}{4}\left(\frac{ V''(f)}{V(f)} \right)\ll1\label{slowroll2} \quad\mbox{,}
\end{equation}
where we  used the notation
\begin{equation}
 f\equiv \frac{\sqrt{8\pi}}{m_{pl}}\phi  \quad\quad \text{ and }\quad\quad   V'(f)\equiv \frac{d V}{d f}\quad\text{.}
\end{equation}

It is possible to associate the comoving wavenumber, $k$, of each mode of the density of perturbations with the inflaton value, $f$, when this mode was leaving the Hubble sphere. We find this using the well known  (e.g. \cite{LiddleLyth}) expression for the number of e-folds, $N(f)$, 
\begin{equation}
 \frac{k}{a_0 H_0}= \exp[60-N(f)] 
\end{equation}
with
\begin{equation}
N(f)= 4 \int^f_{f_e} \frac{V}{V'}\,df\approx (f^2-1/2)\label{efolds} \,\,\text{ .}
\end{equation}

The approximation used in Eq. (\ref{efolds}) is justified by the fact that the assumed deviations of the chaotic inflaton potential ($V\propto f^2$) is very small (the interesting deviations, shown in section \ref{secresults} are of order $5\%$).

The adimensional curvature power spectrum, $\PR$, is related to the inflaton potential, in the slow-roll approximation, by
\begin{equation}
 \PR(k)  \propto \frac{V^3(f(k))}{V'^2(f(k))}\,\mbox{ ,}
\end{equation}
and the normalization of the power spectrum used is the one obtained from WMAP5 \cite{WMAP5}.

From $\PR$ it is possible to obtain $\Pd$, the power spectrum of  the density perturbations using (see e.g. \cite{LiddleLyth})
\begin{equation}
 \Pd(k,t)=\left[\frac{2(1+w)}{5+3w}\right]^2\left(\frac{k}{a(t)\,H(t)}\right)^4 T^2(k)\PR
\label{PRPd}
\end{equation}
where $w$ is the equation of state parameter of the dominant component at instant $t$.

For the transfer function, $T(k)$, we used the analytical fit obtained by \cite{EisensteinHu}, which takes into account the presence of dark energy and baryons.

\section{The mass function}
\label{secmf}
The mass function gives the number density, $n(m)$ of DM halos of mass between $m$ and $m+dm$. In order to calculate it, we need first to obtain the variance from the expression
\begin{equation}
\sigma^2(R,t)=\int_0^\infty W^2(k,R) \Pd(k,t) \,\,\frac{d k}{k}\label{sigma}
\end{equation}
where we adopted a Gaussian window function
\begin{equation}
 W^2(k,R)=\exp(-k^2 R^2)
\end{equation}

In order to evaluate the mass function at the present time, we set
\begin{equation}
 a(t)H(t)=a_0 H_0 = \left(2997.9\right)^{-1} h \mbox{ Mpc}^{-1}\mbox{ . }
\end{equation}

Using the Press-Schechter\cite{PS} formalism, we have
\begin{equation}
n(m) \,dm = -\frac{\bar \rho}{m} \sqrt{\frac{2}{\pi}} \frac{\delta_c(t)}{\sigma^2} \frac{d\sigma}{dm} \exp\left(\frac{-\delta_c^2}{2 \sigma^2}\right)\,dm\quad\mbox{,}
\end{equation}
where $\delta_c\approx 1.61$ is the critical density contrast, $\bar \rho$ is the average density of the universe and the mass, $m$, is related with the length, $R$, from the expression
\begin{equation}
 R(m)= \sqrt{\frac{1}{2\pi}}\left(\frac{m}{\bar \rho}\right)^{\frac{1}{3}}\,\mbox{ .}\label{rm}
\end{equation}

\section{Modified potentials}\label{secpot}
\subsection{Saw-tooth potentials}

The modified potential to which we will refer to  as 'saw-tooth' is constructed by substituting the small field part of the quadratic potential by an oscillatory linear modification. Let $\lambda$ be the wavelength of the oscillation and $f_a$ the field when the modification begins. For $f>f_a$, $V(f)=A f^2$ . For $f<f_a$ we define $f_n=f_a-f\,n\,\lambda$. 

\begin{widetext}
 
For $f_{n}-\frac{1}{4}\lambda< f \leq f_n$, the potential has the form,
\begin{equation}\label{Vsaw1}
 V(f)= A\frac{f_n^2-(1-b)\left(f_n-\frac{1}{4}\lambda\right)^2}{\frac{1}{4}\lambda} +A f_n^2\, \text{ , }
\end{equation}
for $f_{n}-\frac{1}{2}< f \leq f_n-\frac{1}{4}\lambda $,
\begin{equation}	
 V(f)= A\frac{(1-b)\left(f_n-\frac{1}{4}\lambda\right)^2-\left(f_n-\frac{1}{2}\lambda\right)^2}{\frac{1}{4}\lambda}
\left(f-f_n+\frac{1}{4}\lambda\right) +A (1-b)\left(f_n-\frac{1}{4}\lambda\right)^2\text{ , }	
\end{equation}
for $f_{n}-\frac{3}{4}< f \leq f_n-\frac{1}{2}\lambda $,
\begin{equation}	
 V(f)= A\frac{\left(f_n-\frac{1}{2}\lambda\right)^2-(1+b)\left(f_n-\frac{3}{4}\lambda\right)^2}{\frac{1}{4}\lambda}
\left(f-f_n+\frac{1}{2}\lambda\right) +A \left(f_n-\frac{1}{2}\lambda\right)^2	\text{ ,}
\end{equation}
and for $f_{n+1}< f \leq f_n-\frac{3}{4}\lambda $,
\begin{equation}\label{Vsaw4}
 V(f)= A\frac{(1+b)\left(f_n-\frac{1}{4}\lambda\right)^2-\left(f_n-\frac{1}{2}\lambda\right)^2}{\frac{1}{4}\lambda}
\left(f-f_n+\frac{1}{4}\lambda\right) +A (1+b)\left(f_n-\frac{1}{4}\lambda\right)^2	\text{ .}
\end{equation}

\end{widetext}

\begin{figure}[htb]\center
\includegraphics[width=\columnwidth]{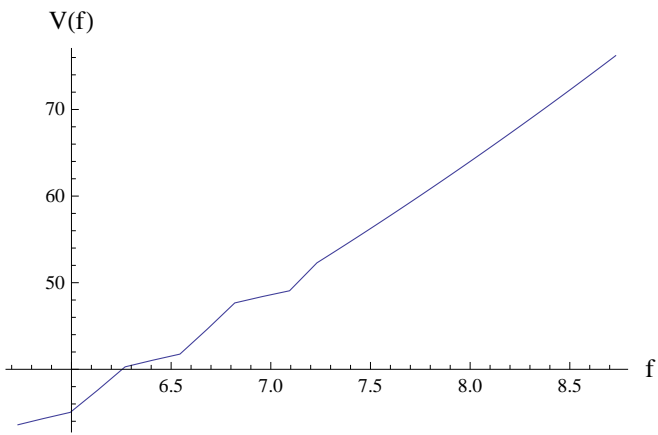}
\caption{Saw-tooth potential, with $b=0.025$ and $\lambda=0.55$}\label{figsaw}
\end{figure}

The potential is shown in Fig. \ref{figsaw}.

\begin{figure}[htb]\center
\includegraphics[width=\columnwidth]{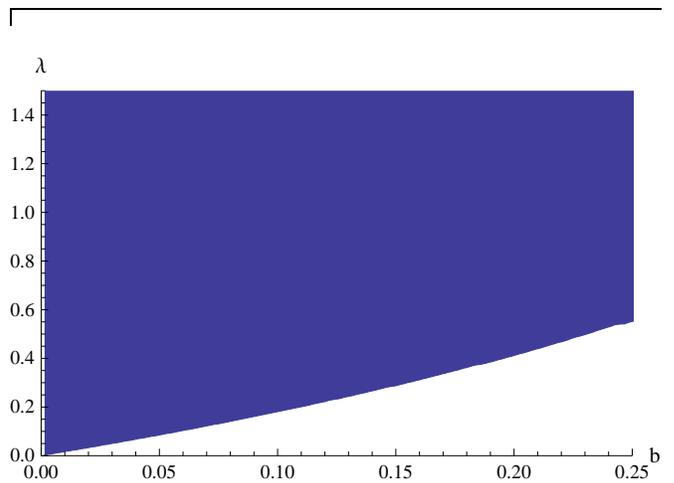}
\caption{The shaded area indicates the region of the parameter space, $(\lambda,b)$, where the saw-tooth potential is compatible with slow-roll inflation.}\label{figparamsaw}
\end{figure}

In Fig. \ref{figsaw} we take $f_a=7.232$ in order to obtain changes in the variance for mass scales $M\lesssim 10^8\msun$. We are in the parameter space region of  slow-roll inflation. The part of the parameter space which allows for this regime -- i.e., where the slow-roll parameters (Eq. (\ref{slowroll1}) and Eq. (\ref{slowroll2})) are smaller than 1 -- is shown in Fig. \ref{figparamsaw}.

\subsection{Sinusoidal potentials}

\begin{figure}[htb]\center
\includegraphics[width=\columnwidth]{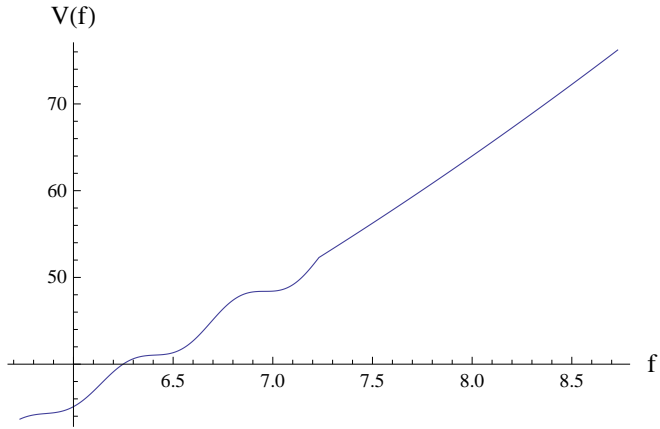}
\caption{Sinusoidal potential, with $b=0.025$ and $\lambda=0.55$}\label{figsin}
\end{figure}

We also analyse the case when the modification of the potential is a simple sine wave,
\begin{equation}
 V(f)= \left\{ 
\begin{array}{l} 
A f^2 \left[1+ b\sin\left(\frac{2\pi\, f}{\lambda} \right) \right]   \,\text{ , }\,  f \leq f_a\\
A f^2    \mbox{ , } f> f_a\\
\end{array}\label{Vsin}
\right. \quad \mbox{,}
\end{equation}
which is plotted in Fig. \ref{figsin}.

\begin{figure}[htb]\center
\includegraphics[width=\columnwidth]{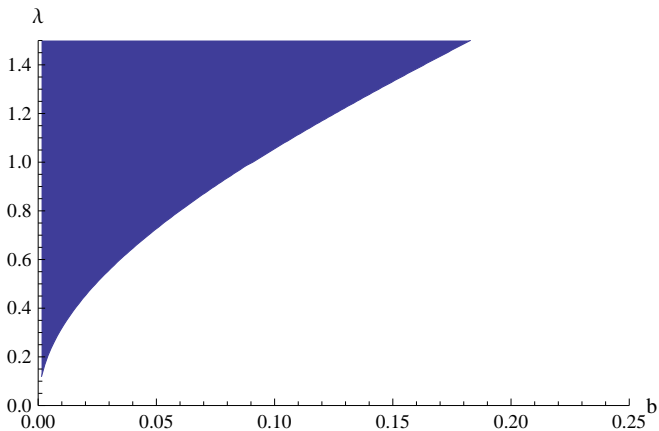}
\caption{The shaded area indicates the region of the parameter space, $(\lambda,b)$, where the sinusoidal potential is compatible with slow-roll inflation.}\label{figparamsin}
\end{figure}

The parameter $f_a$ was chosen using the same criterion as in the saw-tooth case. We tested again for the area of the parameter space which was compatible with the slow-roll regime which is shown in Fig. \ref{figparamsin}.

\section{Results}
\label{secresults}
\begin{figure}[tbp]\center
\includegraphics[width=\columnwidth]{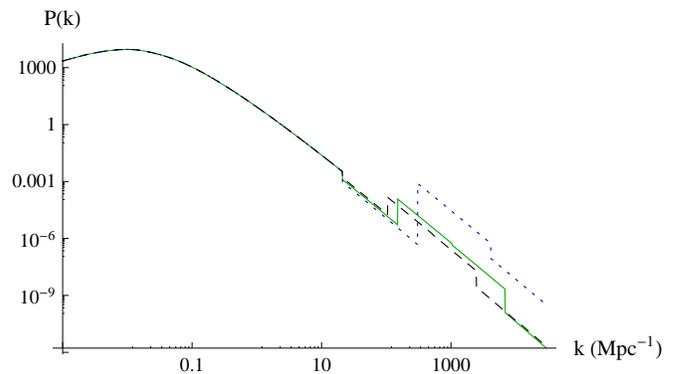}
\caption{Power spectrum obtained using a saw-tooth potential with different values of $b$ and $\lambda$. 
The continuous (\textit{green}) curve corresponds to $b=0.025$ and $\lambda=0.55$;
the dotted (\textit{blue}) curve, to $b=0.050$ and $\lambda=0.75$, and
the dashed (\textit{black}) curve, to $b=0.015$ and $\lambda=0.45$.
}\label{figpksaw}
\end{figure}

We calculated the dimensional density perturbation power spectrum ($P(k)= 2\pi^2 \Pd(k)/ k^3$, following Eq. (\ref{PRPd}), which is shown, for representative parameter values, in Fig. \ref{figpksaw}. 

\begin{figure}[tbp]\center
\includegraphics[width=\columnwidth]{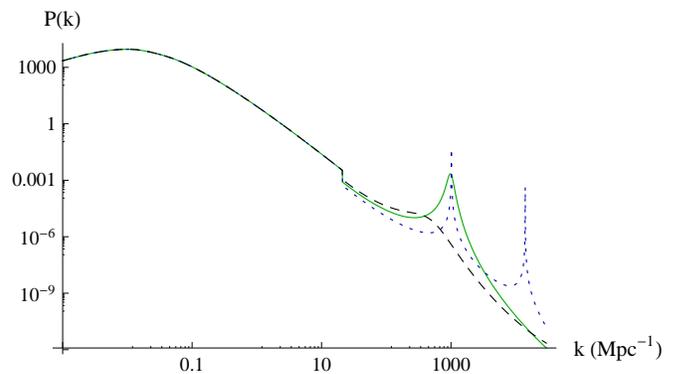}
\caption{Power spectrum obtained using a sinusoidal potential with different values of $b$ and $\lambda$.
The continuous (\textit{green}) curve corresponds to $b=0.025$ and $\lambda=0.55$;
the dotted (\textit{blue}) curve, to $b=0.050$ and $\lambda=0.75$, and
the dashed (\textit{black}) curve, to $b=0.015$ and $\lambda=0.45$.
}\label{figpksin}
\end{figure}

The same quantity was calculated for the sinusoidal potential, as shown in Fig.\ref{figpksin}.

\begin{figure}[tbp]\center 
\includegraphics[width=\columnwidth]{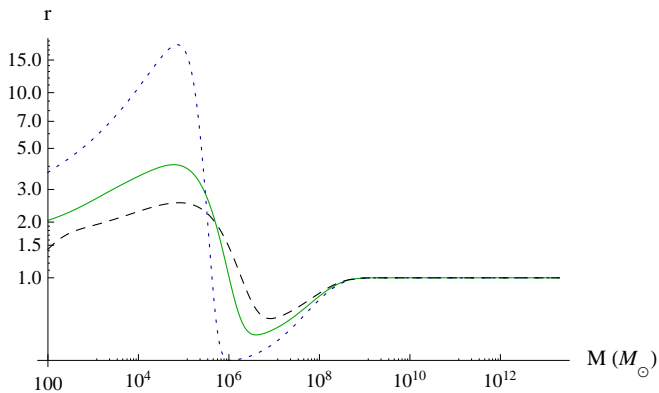}
\caption{Ratio $r=n_{\text{saw-tooth}}(m)/n_{\text{chaotic}}(m)$ .
The continuous (\textit{green}) curve corresponds to $b=0.025$ and $\lambda=0.55$;
the dotted (\textit{blue}) curve, to $b=0.050$ and $\lambda=0.75$, and
the dashed (\textit{black}) curve, to $b=0.015$ and $\lambda=0.45$.
}\label{figratiosaw}
\end{figure}

In Fig. \ref{figratiosaw} we plot the ratio of the mass function obtained from the saw-tooth potential to the mass function of a featureless quadratic potential (i.e., ${n_{saw-tooth}(m)}/{n_{ch}(m)} $ ), for different values of the parameters $b$ and $\lambda$, with $b$ varying from 1.5\% to 5\%. 

When using parameters $b=0.05$ (i.e., a 5\% modification of the chaotic potential) and $\lambda=0.75$, we find a $47\%$ suppression in the number of halos for masses between $10^7$ and $10^8 \msun$.

\begin{figure}[tbp]\center
\includegraphics[width=\columnwidth]{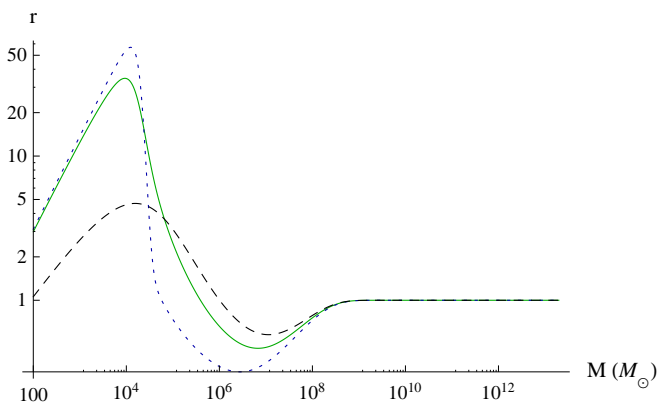}
\caption{Ratio $r=n_{\text{sin}}(m)/n_{\text{chaotic}}(m)$ .
The continuous (\textit{green}) curve corresponds to $b=0.025$ and $\lambda=0.55$;
the dotted (\textit{blue}) curve, to $b=0.050$ and $\lambda=0.75$, and
the dashed (\textit{black}) curve, to $b=0.015$ and $\lambda=0.45$.
}\label{figratiosin}
\end{figure}

In Fig. \ref{figratiosin} we plot the ratio of the mass function obtained from the sinusoidal potential to the mass function of a featureless quadratic potential, again for different values of the parameters $b$ and $\lambda$. Using parameters $b=0.05$ and $\lambda=0.75$, we find a $54\%$ suppression in the number of halos for masses between $10^7$ and $10^8 \msun$.

\section{Conclusions}\label{secconcl}
We  modified the inflationary potential introducing two kinds of oscillatory patterns superimposed, for small field values, on a quadratic potential. Our modifications are small enough to be  compatible with the slow roll conditions and consequently do not alter the usual and successful inflationary predictions for the large scale regime.

The first modification studied (Eq.s (\ref{Vsaw1})-(\ref{Vsaw4})) has the form of a succession of linear segments that creates an oscillating saw-tooth pattern which deviates from the standard quadratic potential by a factor $b$. Using  the  Press-Schechter formalism, we found that the number of small mass halos can be strongly reduced for many choices of the parameters $b$ (which characterizes the amplitude) and $\lambda$ (which characterizes the wavelength). For example, we found that the number of halos with masses between $10^7$ and $10^8\msun$ decreases by 47\% for $b=0.05$ and $\lambda=0.75$.

The second modification studied (Eq. (\ref{Vsin})) is a simple sine function multiplying the quadratic potential. This modification allows one to capture the effects of the previous one without the discontinuities in the second derivatives. Once again we calculated the mass function using the Press-Schechter formalism and  found a strong suppression in the number of small halos for a wide area of the parameter space. As a representative example, we found that the number of halos with masses between $10^7$ and $10^8\msun$ decreases  by 54\% for $b=0.05$ and $\lambda=0.75$.

% We conclude that small oscillatory patterns on the inflaton potential could successfully explain the lack of small galaxies in the Local Group. These oscillatory modifications, however, would leave an important signature in the dynamics of the Local Group: the suppression is accompanied by a significant increase in the number of very small ($\lesssim 10^6\msun$) objects.
We conclude that small oscillatory patterns on the inflaton potential can cause large changes in the predicted halo mass function. In particular, if the oscillations begin in the $10^7-10^8\msun$  range, for example, the oscillations can appreciably suppress the number of dwarf galaxies in this mass range, as observed. It also appreciably increases the number of halos of mass $<10^6\msun$ by factors $\sim 15-60$. This might be a solution to the reionization problem where a very large number of Population \textsc{III} stars in low mass halos are required.

Although we found that a small amplitude ($b=5\%$) saw-tooth oscillatory inflaton potential, with a ``wavelength" $\lambda=0.75$, causes a factor of $\sim 2$ decrease in the number of halos of masses $10^7-10^8 \msun$, present observations indicate a much larger depression -- although the upper limit of the number of observed dwarf galaxies is not well defined due to observational difficulties. From our Fig. \ref{figparamsaw}, however, $\lambda$ could be much less than 0.75 with $b=5\%$, in particular, $\lambda$ could be as small as $\lambda\sim 0.1$, and still be compatible with slow-roll inflation. A detailed analysis of the parameter space $(\lambda,b)$, as well as more general inflation models than the simple single scalar chaotic inflation model which was used here, is thus required (which is now in progress). Only then  can we know whether a small amplitude oscillatory inflaton potential, alone, can resolve the problem of the lack of observed dwarf galaxies in the range $10^7-10^8\msun$.

\section*{Acknowledgments}

L.F.S.R. thanks the Brazilian agency CNPq for financial support (142394/2006-8). 
R.O.thanks the Brazilian agency FAPESP (06/56213-9) and the Brazilian agency CNPq (300414/82-0) for partial
support. 
This research has made use of NASA's Astrophysics Data System.

\bibliographystyle{apsrev}

\begin{thebibliography}{18}
\expandafter\ifx\csname natexlab\endcsname\relax\def\natexlab#1{#1}\fi
\expandafter\ifx\csname bibnamefont\endcsname\relax
  \def\bibnamefont#1{#1}\fi
\expandafter\ifx\csname bibfnamefont\endcsname\relax
  \def\bibfnamefont#1{#1}\fi
\expandafter\ifx\csname citenamefont\endcsname\relax
  \def\citenamefont#1{#1}\fi
\expandafter\ifx\csname url\endcsname\relax
  \def\url#1{\texttt{#1}}\fi
\expandafter\ifx\csname urlprefix\endcsname\relax\def\urlprefix{URL }\fi
\providecommand{\bibinfo}[2]{#2}
\providecommand{\eprint}[2][]{\url{#2}}

\bibitem[{\citenamefont{{Hinshaw} et~al.}(2009)\citenamefont{{Hinshaw},
  {Weiland}, {Hill}, {Odegard}, {Larson}, {Bennett}, {Dunkley}, {Gold},
  {Greason}, {Jarosik} et~al.}}]{WMAP5}
\bibinfo{author}{\bibfnamefont{G.}~\bibnamefont{{Hinshaw}}},
  \bibinfo{author}{\bibfnamefont{J.~L.} \bibnamefont{{Weiland}}},
  \bibinfo{author}{\bibfnamefont{R.~S.} \bibnamefont{{Hill}}},
  \bibinfo{author}{\bibfnamefont{N.}~\bibnamefont{{Odegard}}},
  \bibinfo{author}{\bibfnamefont{D.}~\bibnamefont{{Larson}}},
  \bibinfo{author}{\bibfnamefont{C.~L.} \bibnamefont{{Bennett}}},
  \bibinfo{author}{\bibfnamefont{J.}~\bibnamefont{{Dunkley}}},
  \bibinfo{author}{\bibfnamefont{B.}~\bibnamefont{{Gold}}},
  \bibinfo{author}{\bibfnamefont{M.~R.} \bibnamefont{{Greason}}},
  \bibinfo{author}{\bibfnamefont{N.}~\bibnamefont{{Jarosik}}},
  \bibnamefont{et~al.}, \bibinfo{journal}{The Astrophysical Journals}
  \textbf{\bibinfo{volume}{180}}, \bibinfo{pages}{225} (\bibinfo{year}{2009}),
  \eprint{arXiv:0803.0732}.

\bibitem[{\citenamefont{{The Planck Collaboration}}(2006)}]{planck}
\bibinfo{author}{\bibnamefont{{The Planck Collaboration}}},
  \bibinfo{journal}{ArXiv Astrophysics e-prints}  (\bibinfo{year}{2006}),
  \eprint{arXiv:astro-ph/0604069}.

\bibitem[{\citenamefont{{LSST Science Collaborations: Paul A.~Abell}
  et~al.}(2009)\citenamefont{{LSST Science Collaborations: Paul A.~Abell},
  {Allison}, {Anderson}, {Andrew}, {Angel}, {Armus}, {Arnett}, {Asztalos},
  {Axelrod}, {Bailey} et~al.}}]{LSST}
\bibinfo{author}{\bibnamefont{{LSST Science Collaborations: Paul A.~Abell}}},
  \bibinfo{author}{\bibfnamefont{J.}~\bibnamefont{{Allison}}},
  \bibinfo{author}{\bibfnamefont{S.~F.} \bibnamefont{{Anderson}}},
  \bibinfo{author}{\bibfnamefont{J.~R.} \bibnamefont{{Andrew}}},
  \bibinfo{author}{\bibfnamefont{J.~R.~P.} \bibnamefont{{Angel}}},
  \bibinfo{author}{\bibfnamefont{L.}~\bibnamefont{{Armus}}},
  \bibinfo{author}{\bibfnamefont{D.}~\bibnamefont{{Arnett}}},
  \bibinfo{author}{\bibfnamefont{S.~J.} \bibnamefont{{Asztalos}}},
  \bibinfo{author}{\bibfnamefont{T.~S.} \bibnamefont{{Axelrod}}},
  \bibinfo{author}{\bibfnamefont{S.}~\bibnamefont{{Bailey}}},
  \bibnamefont{et~al.}, \bibinfo{journal}{ArXiv e-prints}
  (\bibinfo{year}{2009}), \eprint{arXiv:0912.0201}.

\bibitem[{\citenamefont{{Pahud} et~al.}(2009)\citenamefont{{Pahud},
  {Kamionkowski}, and {Liddle}}}]{Pahud2008}
\bibinfo{author}{\bibfnamefont{C.}~\bibnamefont{{Pahud}}},
  \bibinfo{author}{\bibfnamefont{M.}~\bibnamefont{{Kamionkowski}}},
  \bibnamefont{and} \bibinfo{author}{\bibfnamefont{A.~R.}
  \bibnamefont{{Liddle}}}, \bibinfo{journal}{Physical Review D}
  \textbf{\bibinfo{volume}{79}}, \bibinfo{pages}{083503}
  (\bibinfo{year}{2009}), \eprint{arXiv:0807.0322}.

\bibitem[{\citenamefont{{Ichiki} and {Nagata}}(2009)}]{Ichiki2009a}
\bibinfo{author}{\bibfnamefont{K.}~\bibnamefont{{Ichiki}}} \bibnamefont{and}
  \bibinfo{author}{\bibfnamefont{R.}~\bibnamefont{{Nagata}}},
  \bibinfo{journal}{Physical Review D} \textbf{\bibinfo{volume}{80}},
  \bibinfo{pages}{083002} (\bibinfo{year}{2009}).

\bibitem[{\citenamefont{{Ichiki} et~al.}(2009)\citenamefont{{Ichiki}, {Nagata},
  and {Yokoyama}}}]{Ichiki2009}
\bibinfo{author}{\bibfnamefont{K.}~\bibnamefont{{Ichiki}}},
  \bibinfo{author}{\bibfnamefont{R.}~\bibnamefont{{Nagata}}}, \bibnamefont{and}
  \bibinfo{author}{\bibfnamefont{J.}~\bibnamefont{{Yokoyama}}},
  \bibinfo{journal}{ArXiv e-prints}  (\bibinfo{year}{2009}),
  \eprint{arXiv:0911.5108}.

\bibitem[{\citenamefont{{Adams} et~al.}(1997)\citenamefont{{Adams}, {Ross}, and
  {Sarkar}}}]{Adams1997}
\bibinfo{author}{\bibfnamefont{J.~A.} \bibnamefont{{Adams}}},
  \bibinfo{author}{\bibfnamefont{G.~G.} \bibnamefont{{Ross}}},
  \bibnamefont{and} \bibinfo{author}{\bibfnamefont{S.}~\bibnamefont{{Sarkar}}},
  \bibinfo{journal}{Nuclear Physics B} \textbf{\bibinfo{volume}{503}},
  \bibinfo{pages}{405} (\bibinfo{year}{1997}), \eprint{arXiv:hep-ph/9704286}.

\bibitem[{\citenamefont{{Brandenberger} and
  {Martin}}(2001)}]{Brandenberger2000}
\bibinfo{author}{\bibfnamefont{R.~H.} \bibnamefont{{Brandenberger}}}
  \bibnamefont{and} \bibinfo{author}{\bibfnamefont{J.}~\bibnamefont{{Martin}}},
  \bibinfo{journal}{Modern Physics Letters A} \textbf{\bibinfo{volume}{16}},
  \bibinfo{pages}{999} (\bibinfo{year}{2001}), \eprint{arXiv:astro-ph/0005432}.

\bibitem[{\citenamefont{{Martin} and {Brandenberger}}(2003)}]{Martin2003}
\bibinfo{author}{\bibfnamefont{J.}~\bibnamefont{{Martin}}} \bibnamefont{and}
  \bibinfo{author}{\bibfnamefont{R.}~\bibnamefont{{Brandenberger}}},
  \bibinfo{journal}{Physical Review D} \textbf{\bibinfo{volume}{68}},
  \bibinfo{pages}{063513} (\bibinfo{year}{2003}),
  \eprint{arXiv:hep-th/0305161}.

\bibitem[{\citenamefont{{Diemand} et~al.}(2007)\citenamefont{{Diemand},
  {Kuhlen}, and {Madau}}}]{ViaLactea}
\bibinfo{author}{\bibfnamefont{J.}~\bibnamefont{{Diemand}}},
  \bibinfo{author}{\bibfnamefont{M.}~\bibnamefont{{Kuhlen}}}, \bibnamefont{and}
  \bibinfo{author}{\bibfnamefont{P.}~\bibnamefont{{Madau}}},
  \bibinfo{journal}{The Astrophysical Journal} \textbf{\bibinfo{volume}{657}},
  \bibinfo{pages}{262} (\bibinfo{year}{2007}), \eprint{arXiv:astro-ph/0611370}.

\bibitem[{\citenamefont{{Springel} et~al.}(2008)\citenamefont{{Springel},
  {Wang}, {Vogelsberger}, {Ludlow}, {Jenkins}, {Helmi}, {Navarro}, {Frenk}, and
  {White}}}]{Aquarius}
\bibinfo{author}{\bibfnamefont{V.}~\bibnamefont{{Springel}}},
  \bibinfo{author}{\bibfnamefont{J.}~\bibnamefont{{Wang}}},
  \bibinfo{author}{\bibfnamefont{M.}~\bibnamefont{{Vogelsberger}}},
  \bibinfo{author}{\bibfnamefont{A.}~\bibnamefont{{Ludlow}}},
  \bibinfo{author}{\bibfnamefont{A.}~\bibnamefont{{Jenkins}}},
  \bibinfo{author}{\bibfnamefont{A.}~\bibnamefont{{Helmi}}},
  \bibinfo{author}{\bibfnamefont{J.~F.} \bibnamefont{{Navarro}}},
  \bibinfo{author}{\bibfnamefont{C.~S.} \bibnamefont{{Frenk}}},
  \bibnamefont{and} \bibinfo{author}{\bibfnamefont{S.~D.~M.}
  \bibnamefont{{White}}}, \bibinfo{journal}{Monthly Notices of the Royal
  Astronomical Society} \textbf{\bibinfo{volume}{391}}, \bibinfo{pages}{1685}
  (\bibinfo{year}{2008}), \eprint{arXiv:0809.0898}.

\bibitem[{\citenamefont{{Koposov} et~al.}(2008)\citenamefont{{Koposov},
  {Belokurov}, {Evans}, {Hewett}, {Irwin}, {Gilmore}, {Zucker}, {Rix},
  {Fellhauer}, {Bell} et~al.}}]{Koposov2008}
\bibinfo{author}{\bibfnamefont{S.}~\bibnamefont{{Koposov}}},
  \bibinfo{author}{\bibfnamefont{V.}~\bibnamefont{{Belokurov}}},
  \bibinfo{author}{\bibfnamefont{N.~W.} \bibnamefont{{Evans}}},
  \bibinfo{author}{\bibfnamefont{P.~C.} \bibnamefont{{Hewett}}},
  \bibinfo{author}{\bibfnamefont{M.~J.} \bibnamefont{{Irwin}}},
  \bibinfo{author}{\bibfnamefont{G.}~\bibnamefont{{Gilmore}}},
  \bibinfo{author}{\bibfnamefont{D.~B.} \bibnamefont{{Zucker}}},
  \bibinfo{author}{\bibfnamefont{H.}~\bibnamefont{{Rix}}},
  \bibinfo{author}{\bibfnamefont{M.}~\bibnamefont{{Fellhauer}}},
  \bibinfo{author}{\bibfnamefont{E.~F.} \bibnamefont{{Bell}}},
  \bibnamefont{et~al.}, \bibinfo{journal}{The Astrophysical Journal}
  \textbf{\bibinfo{volume}{686}}, \bibinfo{pages}{279} (\bibinfo{year}{2008}),
  \eprint{arXiv:0706.2687}.

\bibitem[{\citenamefont{{Macci{\`o}} et~al.}(2010)\citenamefont{{Macci{\`o}},
  {Kang}, {Fontanot}, {Somerville}, {Koposov}, and {Monaco}}}]{Maccio2009}
\bibinfo{author}{\bibfnamefont{A.~V.} \bibnamefont{{Macci{\`o}}}},
  \bibinfo{author}{\bibfnamefont{X.}~\bibnamefont{{Kang}}},
  \bibinfo{author}{\bibfnamefont{F.}~\bibnamefont{{Fontanot}}},
  \bibinfo{author}{\bibfnamefont{R.~S.} \bibnamefont{{Somerville}}},
  \bibinfo{author}{\bibfnamefont{S.}~\bibnamefont{{Koposov}}},
  \bibnamefont{and} \bibinfo{author}{\bibfnamefont{P.}~\bibnamefont{{Monaco}}},
  \bibinfo{journal}{Monthly Notices of the Royal Astronomical Society}
  \textbf{\bibinfo{volume}{402}}, \bibinfo{pages}{1995} (\bibinfo{year}{2010}),
  \eprint{arXiv:0903.4681}.

\bibitem[{\citenamefont{{Koposov} et~al.}(2009)\citenamefont{{Koposov}, {Yoo},
  {Rix}, {Weinberg}, {Macci{\`o}}, and {Escud{\'e}}}}]{Koposov2009}
\bibinfo{author}{\bibfnamefont{S.~E.} \bibnamefont{{Koposov}}},
  \bibinfo{author}{\bibfnamefont{J.}~\bibnamefont{{Yoo}}},
  \bibinfo{author}{\bibfnamefont{H.}~\bibnamefont{{Rix}}},
  \bibinfo{author}{\bibfnamefont{D.~H.} \bibnamefont{{Weinberg}}},
  \bibinfo{author}{\bibfnamefont{A.~V.} \bibnamefont{{Macci{\`o}}}},
  \bibnamefont{and} \bibinfo{author}{\bibfnamefont{J.~M.}
  \bibnamefont{{Escud{\'e}}}}, \bibinfo{journal}{The Astrophysical Journal}
  \textbf{\bibinfo{volume}{696}}, \bibinfo{pages}{2179} (\bibinfo{year}{2009}),
  \eprint{arXiv:0901.2116}.

\bibitem[{\citenamefont{{Baugh}}(2006)}]{Baugh2006}
\bibinfo{author}{\bibfnamefont{C.~M.} \bibnamefont{{Baugh}}},
  \bibinfo{journal}{Reports on Progress in Physics}
  \textbf{\bibinfo{volume}{69}}, \bibinfo{pages}{3101} (\bibinfo{year}{2006}),
  \eprint{arXiv:astro-ph/0610031}.

\bibitem[{\citenamefont{{Liddle} and {Lyth}}(2000)}]{LiddleLyth}
\bibinfo{author}{\bibfnamefont{A.~R.} \bibnamefont{{Liddle}}} \bibnamefont{and}
  \bibinfo{author}{\bibfnamefont{D.~H.} \bibnamefont{{Lyth}}},
  \emph{\bibinfo{title}{{Cosmological Inflation and Large-Scale Structure}}}
  (\bibinfo{year}{2000}).

\bibitem[{\citenamefont{{Eisenstein} and {Hu}}(1998)}]{EisensteinHu}
\bibinfo{author}{\bibfnamefont{D.~J.} \bibnamefont{{Eisenstein}}}
  \bibnamefont{and} \bibinfo{author}{\bibfnamefont{W.}~\bibnamefont{{Hu}}},
  \bibinfo{journal}{The Astrophysical Journal} \textbf{\bibinfo{volume}{496}},
  \bibinfo{pages}{605} (\bibinfo{year}{1998}), \eprint{arXiv:astro-ph/9709112}.

\bibitem[{\citenamefont{{Press} and {Schechter}}(1974)}]{PS}
\bibinfo{author}{\bibfnamefont{W.~H.} \bibnamefont{{Press}}} \bibnamefont{and}
  \bibinfo{author}{\bibfnamefont{P.}~\bibnamefont{{Schechter}}},
  \bibinfo{journal}{The Astrophysical Journal} \textbf{\bibinfo{volume}{187}},
  \bibinfo{pages}{425} (\bibinfo{year}{1974}).

\end{thebibliography}

\end{document}